\documentclass{article}
\usepackage{graphicx} 
\usepackage{physics,mathtools,amsmath,amssymb,amsfonts,latexsym,amscd,euscript,url,hyperref,dsfont,amsthm}
\usepackage{tikz}
\usepackage[a4paper,left=2.5cm,right=2.5cm,top=2.5cm,bottom=2.5cm]{geometry}
\usepackage{mathrsfs}
\usepackage[normalem]{ulem}
\usepackage{cancel}
\usepackage{wrapfig}
\usepackage[affil-it]{authblk}

\usepackage{circuitikz}
\usepackage[sort&compress,numbers]{natbib}
\usepackage{appendix}
\usepackage{comment}

\usepackage{xcolor,ulem}

\newtheoremstyle{named}{}{}{\itshape}{}{\bfseries}{.}{.5em}{\thmnote{#3 }#1}
\theoremstyle{named}
\newtheorem*{namedtheorem}{Theorem}
\usepackage{accents}
\makeatletter
\newcommand{\sbullet}{%
  \hbox{\fontfamily{lmr}\fontsize{.4\dimexpr(\f@size pt)}{0}\selectfont\textbullet}}

\makeatother

\begin{document}
\title{{Gauge aspects of teleparallel gravity and Noether's second theorem}}
\author[1]{Sebastian Brezina%
  \thanks{\texttt{sebastian.brezina@fmph.uniba.sk}}~}
  \author[1,2]{Martin Kr\v{s}\v{s}\'ak%
  \thanks{\texttt{martin.krssak@fmph.uniba.sk}}}
  \affil[1]{Department of Theoretical Physics, Faculty of Mathematics,
Physics and Informatics, Comenius University in Bratislava, 84248, Slovak Republic}
  \affil[2]{Department of Astronomy,  School of Physical Sciences, University of	Science and Technology of China, Hefei, Anhui 230026, China}

\date{}
\maketitle
\begin{abstract}
We examine the gauge structure of the Teleparallel Equivalent of General Relativity (TEGR) from the perspective of Noether's second theorem. We show that diffeomorphisms and local Lorentz transformations are the gauge symmetries of TEGR, as these are the only variational symmetries that give rise to Noether identities. We then discuss the distinct roles of these two symmetries and their implications for the gauge-theoretic interpretation of gravity. Finally, we discuss consequences of these observations for the role of translations in gauge theories of gravity and the coupling of matter in TEGR. In particular, we demonstrate that the matter coupling prescription is uniquely determined and coincides with that of general relativity, including in the case of spinor fields.
\end{abstract}

\tableofcontents

\section{Introduction}

Gauge theories are recognized as a unified framework for describing the fundamental non-gravitational interactions in nature: the electromagnetic interaction, the electroweak interaction, and the strong interaction \cite{Trautman:1970cy,Rudolph:2017pug, Hamilton:2017gbn}. On the other hand, since the 1950s, there has been debate about whether gravitational interaction can also be understood as a gauge theory \cite{Utiyama:1956sy, Kibble:1961ba, Sciama, Trautman:1981fd, Ivanenko:1983fts, Percacci:1984bq, Hehl:1994ue, Blagojevic:2012bc}. This debate is closely related to the more general issue of how gauge theories and their underlying symmetries should be defined, which is essential for developing a consistent framework for describing all four fundamental interactions.

Different definitions of gauge theory and gauge symmetry lead to different conclusions about the status of gravity as a gauge theory. Some definitions do not allow gravity to be interpreted as a gauge theory due to their inability to incorporate diffeomorphism symmetry, which is the fundamental symmetry of gravity theories. Let us illustrate this with the following definitions of gauge transformations.

If one adopts the geometric definition of gauge theory commonly used in the mathematical formulation of gauge theories of internal symmetries, such as Yang–Mills theory \cite{Yang:1954ek,Utiyama:1956sy}, gauge transformations are understood as vertical automorphisms of a principal $G$-bundle \cite{Husemoller, bleecker1981gauge,Mangiarotti:2000ce,Rudolph:2017pug, Hamilton:2017gbn},  then no gravitational theory can be viewed as a gauge theory, since diffeomorphisms of the spacetime manifold $M$ cannot be described by vertical automorphisms. On the other hand, if one uses, for example, Trautman's framework \cite{Trautman:1979fg, Trautman-G, Trautman:1981fd, Brezina:2026nxx}, in which gauge transformations are defined as automorphisms of a principal $G$-bundle that preserve the absolute elements of the theory \cite{anderson1967principles, Giulini:2006yg}, then both Yang–Mills theory and gravity can be viewed as gauge theories within this framework.

The original approach to gauge theory, commonly found in the physics literature and which motivated the development of Yang-Mills theory \cite{Yang:1954ek}, is through Weyl's gauge principle, where global symmetries are promoted to local symmetries \cite{Weyl:1918ib, Weyl:1918pdp, Weyl:1929fm}. In this approach, gauge transformations are understood as maps of open sets $U\subset M$ to the group $G$  \cite{bleecker1981gauge, Trautman:1979fg, Hamilton:2017gbn}. However, in the fiber bundle language, the group $G$ is the structure group of the principal bundle, and local symmetries obtained in this way correspond only to vertical automorphisms, which again cannot accommodate diffeomorphisms \cite{Trautman:1970cy, Ivanenko:1983fts, Sardanashvily:1984dx, Ivanenko:1987ka}.

The need to incorporate diffeomorphisms and to consider them gauge symmetries is transparent from the perspective of Noether's second theorem, one of the two celebrated theorems formulated by Emmy Noether in 1918 \cite{Noether:1918zz}.  The first theorem concerns infinitesimal variational symmetries of the action that depend on a finite number of parameters, and yields on-shell conservation laws. The second theorem, by contrast, applies to infinitesimal variational symmetries that are linearly parametrized by arbitrary functions of spacetime and their derivatives, and yields off-shell differential identities among the Euler-Lagrange expressions, known as the Noether identities.  We adopt precisely this second class of symmetries as our definition of gauge transformations \cite{Henneaux:1992ig, Fulp:2002fm, Avery:2015rga, anderson1967principles, kosmann2011noether, Torre2022}. 
From this perspective, both the internal gauge symmetries of Yang–Mills theories and diffeomorphisms in gravitational theories are gauge symmetries, since both give rise to Noether identities. In the case of diffeomorphisms in general relativity, the relevant Noether identity is the (contracted) Bianchi identity\footnote{The Bianchi identities played an important role in discovering Noether's theorems, and understanding the problem of energy in general relativity, see  \cite{Noether:1918zz} and \cite{DeHaro:2021gdv}. This is why the Noether identities are also known as the generalized Bianchi identities. For historical context of Noether's theorems,  we recommend reading \cite{kosmann2011noether}.}. This view is also consistent with the Hamiltonian formulation and Dirac-Bergmann constraint analysis \cite{Dirac1950, Bergman1951}, in which gauge transformations are constructed from the first-class constraints \cite{Henneaux:1992ig}.

The Teleparallel Equivalent of General Relativity (TEGR) is a classical gravitational theory based on teleparallel geometry that is dynamically equivalent to general relativity \cite{Aldrovandi:2013wha,Krssak:2018ywd,Maluf:2003fs}.  Originally introduced by Einstein in the 1920s as an unsuccessful attempt for unified theory of gravity and electromagnetism \cite{Einstein1929a,Sauer:2004hj,Goenner:2004se,Krssak:2024xeh},  it was later revisited by Møller \cite{Moller1961}, who showed that it can be formulated as a consistent theory of gravity only,  dynamically equivalent to general relativity. Since the 1970s, TEGR has often been presented as a classical gauge gravitational theory of translations; see, for example, \cite{Hayashi:1967se, Cho:1975dh, Aldrovandi:2013wha, Pereira:2019woq, Krssak:2018ywd, CANTATA:2021asi}.

The first goal of this work is to analyze the gauge structure of TEGR through Noether's second theorem. Throughout the analysis, we construct a model of a classical gauge theory of translations to demonstrate crucial structural differences between this theory and TEGR. The provided analysis results in the conclusion that the translational gauge symmetry is not present among the gauge symmetries of TEGR, which are local Lorentz transformations and (orientation-preserving) diffeomorphisms of $M$. This conclusion is consistent with the conclusion in our previous work \cite{Brezina:2025dbc}, where the analysis of the translational gauge interpretation of TEGR was carried out using the theory of Ehresmann connections on principal $G$-bundles \cite{KN1} and Trautman's framework for classical gauge theories \cite{Trautman:1979fg, Trautman-G, Trautman:1981fd}. We concluded, in agreement with \cite{Fontanini:2018krt, LeDelliou:2019esi}, and with \cite{Sardanashvily:1983xbn, Sardanashvily:1984dx, Ivanenko:1983fts, Ivanenko:1987ka, Dass:1984qk} in the more general context of Poincaré gauge theories, that the interpretation of TEGR as a classical gauge theory of translations, i.e., using principal $T(4)$-bundle endowed with a translational connection, is problematic due to identification of a translational gauge field with an orthonormal frame field. We explained that the formulation with the Lorentz group (or the equivalent formulation with the Poincaré group) as the structure group of the theory is consistent, and the gauge transformations, in the sense of Trautman, are diffeomorphisms of $M$.

The second goal of this work is to clarify the situation with matter coupling in TEGR, which has been discussed in the literature for a long time \cite{Hayashi:1979qx,deAndrade:2001vx,Obukhov:2002tm,Mielke:2004gg,Maluf:2003fs,Obukhov:2004hv,BeltranJimenez:2020sih,Hohmann:2022mlc,Ferraro:2022xwk,Aldrovandi:2013wha,Blagojevic:2012bc}. These discussions conclude that the teleparallel coupling leads to inconsistencies, while the Levi-Civita coupling is the consistent choice. Therefore, the Levi-Civita coupling is postulated for consistency reasons. We argue that the structure of Noether identities in TEGR directly implies the Levi-Civita coupling. Thus, the Levi-Civita coupling in TEGR is not postulated just for consistency reasons, which are a basic requirement on any physical theory, but is implied by the structure of TEGR itself. On the other hand, in modified theories, such as new general relativity \cite{Hayashi:1979qx} or $f(T)$ \cite{Ferraro:2006jd, Ferraro:2008ey,Linder:2010py}, one has the possibility to consider both the teleparallel and the Levi-Civita couplings. Both options are consistent; thus, one of them has to be postulated.

\section{Noether's theorems and their applications}\label{sec0}
In this part, we focus on Noether's theorems and several basic applications of Noether's second theorem in gauge theories of internal symmetries over Minkowski spacetime and in metric-affine gauge gravitational theories and their subclass Lorentz gauge theories. We start with recalling several basic facts about actions, Lagrangians, and their infinitesimal variations in the language of differential forms. While all the results of this section are known, \cite{Hehl:1994ue,Obukhov:2006fy,Obukhov:2006ge,Obukhov:2006sk,Obukhov:2007sh}, we consider it important to summarize and discuss the results for the overall logic and understanding of the concepts, in support of Section \ref{sec3}. We assume that all structures, maps, and geometric objects used here are smooth. We follow definitions and conventions used in \cite{Fecko:2006zy}. 

\subsection{Classical Lagrangian field theory and Noether's theorems}\label{sec1}  

Let $M$ be a $n$-dimensional manifold and $\psi^a$, where $a=\{1,...,k\}$, be differential forms of degrees $p_a$ representing fields on $M$. We assume that the studied system is variational and its  \emph{Lagrangian $n$-form} $L$ depends on fields $\psi$ and their first derivatives $d\psi$. The action over a fixed domain $U\subset M$ is 
\begin{equation}\label{sec1-1}
    \mathcal{S}[\psi]=\int_UL(\psi,d\psi).
\end{equation}
An infinitesimal variation of the Lagrangian $n$-form is given as
\begin{equation}\label{sec1-2}
    \delta L=E_a\wedge\delta\psi^a+d\Theta,
\end{equation}
where the $\left(n-p_a\right)$-forms $E_a$ are the \emph{Euler-Lagrange expressions} and the $(n-1)$-form $\Theta:=\Theta(\psi,\delta\psi)$ is linear in $\delta\psi$. The corresponding variation of the action \eqref{sec1-1} is 
\begin{equation}\label{sec1-3}
    \delta\mathcal{S}=\int_UE_a\wedge\delta\psi^a+\int_{\partial U}\Theta.
\end{equation}
To derive the Euler-Lagrange equations, i.e, the field equations of the fields $\psi^a$, we require $\delta\mathcal{S}=0$ and the variations $\delta \psi^a$ to have compact support in $U$ -- one can also require a weaker condition $\left(\delta \psi^a\right)|_{\partial U}=0$. Then by using the fundamental lemma of variational calculus, one obtains the Euler-Lagrange equations
\begin{equation}\label{sec1-4}
    E_a=0.
\end{equation}

We call the variation $\delta\psi^a$ an (infinitesimal variational) \emph{symmetry} of the Lagrangian $n$-form $L$ iff $\delta L=dK$, where $K$ is a $(n-1)$-form linear in $\delta\psi^a$. We say that the action $\mathcal{S}[\psi]$ is \emph{quasi-invariant} under the variation $\delta\psi^a$ iff $\delta \mathcal{S}=\int_{\partial U}K$, and invariant iff $\delta\mathcal{S}=0$\footnote{While the original Noether's theorems considered the strict invariance of actions, they were generalized to quasi-invariance, i.e., to invariance up to a boundary term,  by Bessel-Hagen \cite{Bessel-Hagen1921}.}.

Let us assume that the infinitesimal variation of the fields $\delta\psi^a$ is linearly parametrized by a set of $\rho$ parameters $\xi^i$, where $i=\{1,...,\rho\}$, i.e.,
\begin{equation}\label{sec1-5}
    \delta_{\xi}\psi^a:=R_i^a(\psi,d\psi)\xi^i,
\end{equation}
where $R^a_i(\psi,d\psi)$ are $p_a$-forms. The variation \eqref{sec1-5} corresponds to the infinitesimal action of a one-parameter subgroup of an $\rho$-dimensional Lie group $G_{\rho}$ on fields $\psi^a$. Noether's first theorem states
\begin{namedtheorem}[Noether's First]\label{firsttheorem}
If the action \eqref{sec1-1} is quasi-invariant with respect to an $\rho$-dim.~Lie group $G_\rho$, then there are $\rho$ linearly independent combinations of the Euler-Lagrange expressions that are exact. 
\end{namedtheorem}
These (off-shell) relations can be obtained using \eqref{sec1-3}, \eqref{sec1-5}, and the arbitrariness of $\xi^i$ and $U$. They read
\begin{equation}\label{sec1-6}
    E_a\wedge R^a_i=dJ_i,
\end{equation}
where the $(n-1)$-forms $J_i:=K_i-\Theta_i$ are the \emph{Noether currents}, where $K_i:=K\left(\psi,R_i\right)$, $\Theta_i:=\Theta\left(\psi,R_i\right)$. From \eqref{sec1-6} follows that the Noether currents are conserved on-shell, i.e., when the Euler-Lagrange equations are satisfied by the fields $\psi^a$ we get $dJ_i\approx0$, where $\approx$ means on the solutions of $E_a=0$. 

The converse of Noether's first theorem is also true, and the theorem remains valid in the limiting case of an infinite number of parameters. To clarify the terminology, the transformations \eqref{sec1-5} are often referred to as rigid or global symmetries.

Let us now assume that the infinitesimal variation of the fields $\delta\psi^a$ is linearly parametrized by a set of $\rho$ arbitrary functions $\lambda^i(x)$ and their first derivatives\footnote{In the original work of Noether, this parametrization is assumed to be linear up to a finite order of derivatives of $\lambda^i(x)$.}, where $i=\{1,...,\rho\}$, i.e., 
\begin{equation}\label{sec1-7}
    \delta_{\lambda}\psi^a:=R_i^a(\psi,d\psi)\lambda^i(x)+S^a_i(\psi,d\psi)\wedge d\lambda^i(x),
\end{equation}
where $R^a_i(\psi,d\psi)$ are $p_a$-forms and $S^a_i(\psi,d\psi)$ are $\left(p_a-1\right)$-forms. The variation \eqref{sec1-7} corresponds to the infinitesimal action of a one-parameter subgroup of an infinite-dim.~group $G_{\infty \rho}$ on fields $\psi^a$.
 
An \emph{infinitesimal gauge transformation} is defined as the infinitesimal variation \eqref{sec1-7}. The infinite-dimensional group $G_{\infty \rho}$ is called the \emph{gauge group}. An infinitesimal gauge transformation is an (infinitesimal) \emph{gauge symmetry} of the Lagrangian $n$-form iff $\delta_{\lambda} L=dK$ under variation \eqref{sec1-7}. Then we say that the action $\mathcal{S}[\psi]$ has a \emph{gauge symmetry} iff $\delta_{\lambda} \mathcal{S}=\int_{\partial U}K$ under variation \eqref{sec1-7}.

Noether's second theorem states
\begin{namedtheorem}[Noether's Second]\label{secondtheorem}
If the action \eqref{sec1-1} is quasi-invariant under the infinite-dimensional group $G_{\infty \rho}$, then there are $\rho$ identities among the Euler-Lagrange expressions and their first derivatives.
\end{namedtheorem}
These (off-shell) identities, called \emph{Noether identities}, can be obtained using \eqref{sec1-3}, \eqref{sec1-7}, the fundamental lemma of variational calculus, and assuming compactly supported $\lambda^i(x)$ in a fixed $U$. The identities are
\begin{equation}\label{sec1-8}
    E_a\wedge R^a_i+(-1)^nd\left( E_a\wedge S^a_i\right)=0,
\end{equation}
The identities \eqref{sec1-8} tell us that not all Euler-Lagrange expressions are independent in the presence of gauge symmetries. The converse of Noether's second theorem is also true. Therefore, the existence of gauge symmetries of an action is related to the existence of Noether identities.

\subsection{Electromagnetism and Yang-Mills theory}\label{subsec2-1} 
Let us consider actions defined in Minkowski spacetime $\left(\mathbb{R}^4,\eta,\nabla_{LC}\right)$, where $\eta$ is the Minkowski metric tensor field with the signature $(+~-~-~-)$ and $\nabla_{LC}$ is the Levi-Civita connection.

\emph{Electromagnetism} without sources is described by the action 
\begin{equation}\label{subsec2-1-1}
    \mathcal{S}[A]:=-\frac{1}{2}\int_{U}F\wedge*F,
\end{equation}
where $A$ is a 1-form of electromagnetic potential (gauge field), $F:=dA$ is the 2-form of electromagnetic field strength, and $*$ is the Hodge star depending on $\eta$ and orientation, which we choose to be right-handed.
The Euler-Lagrange expression of \eqref{subsec2-1-1} is
\begin{equation}\label{subsec2-1-2}
    E=d*F.
\end{equation}
The action \eqref{subsec2-1-1} has several continuous symmetries. Namely, it is invariant under the \emph{Poincaré transformations}\footnote{More generally, the action \eqref{subsec2-1-1} is invariant under conformal transformations constituting the group $SO(2,4)$.} constituting the Poincaré group $\mathcal{P}=SO(1,3)\ltimes T(4)$ and \emph{local $U(1)$ transformations}, i.e, under the (smooth) maps $S:U\rightarrow U(1)$ that assign to a point in an open set $U$ an element of $U(1)$ group. The Poincaré group is a finite-dim.~Lie group and leads to on-shell conservation laws through Noether's first theorem. Local $U(1)$ transformations constitute the infinite-dim.~gauge group $G_{\infty1}$, induce infinitesimal variations $\delta_{\lambda} A=d\lambda(x)$, and lead through  Noether's second theorem to the Noether identity
\begin{equation}\label{subsec2-1-3}
    dE=0.
\end{equation}

The electromagnetism case can be generalized to Yang-Mills theory. Let $G$ be a semi-simple compact $\rho$-dim.~Lie group with the Lie algebra $\mathcal{G}$, $\mathcal{A}$ be a $\mathcal{G}$-valued 1-form, called gauge field, $\mathcal{F}:=\mathcal{D}\mathcal{A}$ its gauge field strength, which is a $\mathcal{G}$-valued 2-form, and $\mathcal{D}$ denotes the exterior covariant derivative on the base \cite[p.~604]{Fecko:2006zy}. The action of \emph{Yang-Mills theory} without sources is given by 
\begin{equation}\label{subsec2-1-6}
    \mathcal{S}[\mathcal{A}]:=-\frac{1}{2}\int_U k_{ij}\mathcal{F}^i\wedge*\mathcal{F}^j,
\end{equation}
where $k_{ij}$ is the Killing-Cartan form. A variation of this action yields the Euler-Lagrange expression
\begin{equation}\label{subsec2-1-7}
    E=\mathcal{D}*\mathcal{F}.
\end{equation}
The action \eqref{subsec2-1-6} also has several continuous symmetries. It is invariant under the Poincaré transformations (more generally, under the conformal transformations) and local $G$ transformations, i.e., under the (smooth) maps $S:U\rightarrow G$ that assign to a point in an open set $U$ an element of $G$ group. The Poincaré group leads to on-shell conservation laws through Noether's first theorem. Local $G$ transformations constitute the infinite-dim.~gauge group $G_{\infty\rho}$, induce infinitesimal variations $\delta_{\lambda} A=\mathcal{D}\lambda(x)$, where $\lambda(x)$ is a $\mathcal{G}$-valued 0-form, and lead through Noether's second theorem to the Noether identity
\begin{equation}\label{subsec2-1-8}
    \mathcal{D}E=0.
\end{equation}

Based on the results \eqref{subsec2-1-3} and \eqref{subsec2-1-8}, which can be obtained by applying $d$ to Maxwell's $E$ \eqref{subsec2-1-2} and $\mathcal{D}$ to the Yang-Mills $E$ \eqref{subsec2-1-7}, one may naturally think that Noether's second theorem does not provide any useful insight into the symmetry structure of \eqref{subsec2-1-1} and \eqref{subsec2-1-6}. However, here one has to realize that the real power of Noether's second theorem manifests when the action is not given explicitly, as is in the considered cases, but when one requires a generic action depending on the fields of interest to admit invariance under some given gauge group $G_{\infty\rho}$. For example, consider an action $\mathcal{S}[A]$ and assume that the gauge group $G_{\infty1}$ is the same as in the case of \eqref{subsec2-1-1}. Then Noether's second theorem puts restrictions on the possible forms of the Euler-Lagrange expression $E$ of $\mathcal{S}[A]$, i.e., on the Lagrangian 4-form $L(A,dA)$. The Euler-Lagrange expression $E$ has to satisfy Noether identity \eqref{subsec2-1-3}. The choice $L(A,dA)=F\wedge*F$ is one of several; another is, for example, $L(A,dA)=F\wedge F$ with trivial Euler-Lagrange expression $E=0$.

Furthermore, if a source-free action is given and one wants to incorporate external sources, then Noether's second theorem puts non-trivial restrictions on these sources to preserve the consistency of the field equations. We illustrate this fact using the example of electromagnetism in the presence of an external background source. Let us assume that the 3-form of external background current $J$, which represents the source of the electromagnetic field, is given and that the action has the form
\begin{equation}\label{subsec2-1-4}
    \mathcal{S}[A]:=-\frac{1}{2}\int_UF\wedge*F-\int_UA\wedge J.
\end{equation}
The Euler-Lagrange expression and the Noether identity are
\begin{equation}\label{subsec2-1-5}
  E=d*F+J,\qquad dE=0.
\end{equation}
The Noether identity \eqref{subsec2-1-5} yields $dJ=0$ off-shell. This result imposes a consistency requirement on the coupling of external background currents to the electromagnetic field, stating that only off-shell conserved external background currents yield consistent coupling. 

\subsection{Metric-affine gauge theory and Lorentz gauge theory}\label{subsec2-2}
We start with the \emph{metric-affine gauge gravitational theories}\footnote{\label{footnote1}It would be more appropriate to use the term "linear" instead of "affine" since in these theories one practically uses a linear connection on the frame bundle, whose structure group is the general linear group. Affine connections are defined on the affine principal bundle, whose structure group is the affine group. A well-known theorem \cite[p.~129]{KN1} states that there is a one-to-one correspondence between affine and linear connections. This explains the use of the term "affine".}, a class of gravitational theories based on the geometric structure $\left(M,g,\nabla\right)$, where $M$ is a  4-dim.~orientable manifold, $g$ is a metric tensor field with the signature $(+~-~-~-)$, and $\nabla$ is a (general) linear connection. A linear connection is locally described by linear connection $1$-forms $\omega=\omega^a{}_bB^b{}_a$, where $B^b{}_a$ is a basis in the $gl(4,\mathbb{R})$ Lie algebra, and is characterized by three tensor fields: the \emph{curvature $2$-forms} $\Omega^a{}_b$, the \emph{torsion $2$-forms} $T^a$, and the \emph{non-metricity $1$-forms} $Q_{ab}$ defined as
\begin{equation}\label{subsec2-2-1}
    \Omega^a{}_b:=\mathcal{D}\omega^a{}_b,\qquad T^a:=\mathcal{D}e^a,\qquad Q_{ab}:=\mathcal{D}g_{ab},
\end{equation}
where $e^a$ is a coframe field, $g_{ab}$ are the components of the metric tensor field $g$ with respect to the frame field $e_a$ dual to $e^a$, and $\mathcal{D}$ is the exterior covariant derivative on the base. The frame indices are raised/lowered using $g_{ab}$, e.g., $\Omega^{ab}=g^{bc} \Omega^a{}_c $ or $\Omega_{ab}=g_{ac} \Omega^c{}_b $.

We assume that the actions of the considered gravitational theories are given as
\begin{equation}\label{subsec2-2-2}
   \mathcal{S}\left[e^a,\omega^a{}_b,g_{ab}\right]:=\int_{U}L\left(e^a,\omega^a{}_b,g_{ab},de^a,d\omega^a{}_b,dg_{ab}\right),
\end{equation}
whose infinitesimal variation is
\begin{equation}\label{subsec2-2-3}
    \delta \mathcal{S}=\int_{U}\left(E_a\wedge\delta e^a+E^b{}_a\wedge\delta \omega^a{}_b+\Tilde{E}^{ab}\wedge\delta g_{ab}\right)+\int_{\partial U}\Theta.
\end{equation}
Since $L$ in \eqref{subsec2-2-2} is constructed tensorially from the fields $\left\{e^a,\omega^a{}_b,g_{ab},de^a,d\omega^a{}_b,dg_{ab}\right\}$, the action \eqref{subsec2-2-2} is trivially invariant under the \emph{local $GL(4,\mathbb{R})$ transformations}, i.e., under the (smooth) maps $S:U\rightarrow GL(4,\mathbb{R})$ that assign to a point of $U\subset M$ an element of $GL(4,\mathbb{R})$ group. Moreover, the action is assumed to be invariant under the \emph{diffeomorphisms} of $M$ constituting the group Diff~$M$. These continuous symmetries of the action are gauge symmetries in the sense of Noether's second theorem, which can be parametrized as in \eqref{sec1-7}, and hence lead to  Noether identities.

Let us first derive the Noether identity for local $GL(4,\mathbb{R})$ gauge symmetry. The fields $g_{ab},~e^a,~\text{and}~\omega^a{}_b$ transform  under a local $GL(4,\mathbb{R})$ transformation given by matrix $A^a{}_b(x)$ as
\begin{equation}\label{subsec2-2-4}
   e'^a=(A^{-1})^a{}_b(x)e^b,\quad \omega'^a_{~b}=(A^{-1})^a{}_c(x)\omega^c{}_dA^d{}_b(x)+(A^{-1})^a{}_c(x)dA^c{}_b(x),\quad g'_{ab}=A^a{}_c(x)A^d{}_b(x)g_{cd},
\end{equation}
which, in the infinitesimal case of a one-parameter subgroup of local $GL(4,\mathbb{R})$, i.e., in the case $A^a{}_b(x)=\delta^a_b+\varepsilon\lambda^a{}_b(x)$, yields infinitesimal variations $\left(\lambda_{(ab)}(x):=g_{c(a}\lambda^c{}_{b)}(x)\right)$
\begin{equation}\label{subsec2-2-5}
  \delta_{\lambda} e^a=-\lambda^a{}_b(x)e^b,\qquad\delta_{\lambda}\omega^a{}_b=\mathcal{D}\lambda^a{}_b(x),\qquad \delta_{\lambda} g_{ab}=2\lambda_{(ab)}(x).
\end{equation}
These variations are of the form \eqref{sec1-7} which means that through \eqref{subsec2-2-3}, invariance of \eqref{subsec2-2-2} under \eqref{subsec2-2-4} and compactly supported $\lambda^a{}_b(x)$ in a fixed $U$ one gets the \emph{Noether identity for local $GL(4,\mathbb{R})$ gauge symmetry}
\begin{equation}\label{subsec2-2-6}
 \mathcal{D}E^a{}_b-E_b\wedge e^a+2g_{bc}\Tilde{E}^{ac}=0.
\end{equation}

The action \eqref{sec1-1} is said to be invariant under diffeomorphisms $f$ of $M$ if and only if it satisfies\footnote{\label{footnote3}In electromagnetism \eqref{subsec2-1-1} or Yang-Mills theories \eqref{subsec2-1-6}, the actions are not invariant under all diffeomorphisms $f$ since the Minkowski metric tensor and right-hand orientation are fixed, thus the requirement \eqref{subsec2-2-7} on possible diffeomorphisms $f$ is to satisfy $f^*\eta=\eta$ and to preserve the orientation. This yields just the Poincaré transformations.}
\begin{equation}\label{subsec2-2-7}
    \int_{f(U)}L\left[\left(f^{-1}\right)^*\psi\right]=\int_{U}L[\psi],
\end{equation}
where $L[\psi]:=L(\psi,d\psi)$. From arbitrariness of $U$ follows the requirement on $L$
\begin{equation}\label{subsec2-2-8}
   f^*\left(L[\psi]\right)=L\left[f^*\psi\right].
\end{equation}
For the infinitesimal action of the one-parameter subgroup $\Phi_t$ of Diff~$M$ generated by a vector field $V$ on fields $\psi$, i.e., for $\Phi_{\varepsilon}^*\psi=\psi+\varepsilon \mathcal{L}_V\psi$, the infinitesimal variation of $\psi$ becomes $\delta_V\psi=\mathcal{L}_V\psi$ and \eqref{subsec2-2-8} yields $\delta_V L=\mathcal{L}_VL$. Employing the Cartan formula $\mathcal{L}_V=di_V+i_Vd$, where $i_V$ is the interior product, and using the fact that $dL=0$, we get for a fixed $U$
\begin{equation}\label{subsec2-2-9}
  \delta_V\mathcal{S}=\int_{\partial U}i_VL.
\end{equation}
This means that diffeomorphism invariance of the action $\mathcal{S}[\psi]$ is a gauge symmetry since infinitesimal variations of $\psi$ are of the form \eqref{sec1-7} and leave the action \eqref{subsec2-2-9} quasi-invariant.

Let us now derive the Noether identity for diffeomorphism gauge symmetry in the case of action \eqref{subsec2-2-2}. The infinitesimal variations of fields $g_{ab},~e^a,~\text{and}~\omega^a{}_b$ induced by  an infinitesimal diffeomorphism $\Phi_{\varepsilon}$ are 
\begin{equation}\label{subsec2-2-10}
  \delta_{V}g_{ab}=\mathcal{L}_Vg_{ab},\qquad \delta_{V} e^a=\mathcal{L}_Ve^a,\qquad \delta_{V}\omega^a{}_b=\mathcal{L}_V\omega^a{}_b.
\end{equation}
Through \eqref{subsec2-2-3}, invariance of \eqref{subsec2-2-2} under \eqref{subsec2-2-10}, employment of definitions \eqref{subsec2-2-1} and Cartan formula in \eqref{subsec2-2-10}, the use of \eqref{subsec2-2-6}, and requiring that the components $V^a(x)$ of $V$ being compactly supported in some fixed $U$ one gets the \emph{Noether identity for diffeomorphism gauge symmetry}
\begin{equation}\label{subsec2-2-11}
 \mathcal{D}E_a+E_c\wedge i_aT^c+E^b{}_c\wedge i_a\Omega^c{}_b +\Tilde{E}^{bc}i_aQ_{bc}=0.
\end{equation}

We continue with the so-called \emph{Lorentz gauge theories}\footnote{\label{footnote2}These theories are also known under the name \emph{Poincaré gauge theories}, in which one works with the Poincaré connections on Poincaré bundles, which are subbundles of the affine bundle. A Poincaré connection is related to a metric linear connection on an orthonormal bundle $OM$, which is a subbundle of the frame bundle $LM$. From the well-known theorem, see footnote \ref{footnote1}, it follows that Poincaré and Lorentz gauge theories are equivalent theories; despite this, the Lorentz gauge theory is usually treated as a special case of the Poincaré one.} which are gravitational theories geometrically described as $\left(M,g,\nabla_m\right)$, where $M$ is a smooth 4-dim.~right-oriented manifold, $g$ is a metric tensor field with the signature $(+~-~-~-)$, and $\nabla_m$ a metric linear connection, i.e., its non-metricity 1-forms $Q_{ab}$ vanish. Furthermore, it is assumed that $g_{ab}=\eta_{ab}$, thus, we restrict to orthonormal frame fields $e_{a}$. Then the metric tensor field is given as $g=\eta_{ab}e^a\otimes e^b$ and the metric linear connection 1-forms $\omega_{ab}$ are antisymmetric, i.e., $\omega_{ab}=-\omega_{ba}$. A generic action is of the form (except special cases, such as general relativity, whose Lagrangian depends tensorially just on $e^a$ and its first and second derivatives)
\begin{equation}\label{subsec2-2-11}
   \mathcal{S}\left[e^a,\omega^a{}_b\right]:=\int_{U}L\left(e^a,\omega^a{}_b,de^a,d\omega^a{}_b\right),
\end{equation}
where $L$ is tensorially constructed from the fields $\left\{e^a,\omega^a{}_b,de^a,d\omega^a{}_b\right\}$. Thus, the action \eqref{subsec2-2-11} is \emph{locally Lorentz invariant}, i.e., invariant under the (smooth) maps $S:U\rightarrow SO(1,3)$ that assign to a point in $U\subset M$ an element of $SO(1,3)$ group. Moreover, the action \eqref{subsec2-2-11} is also assumed to be \emph{diffeomorphism invariant} \eqref{subsec2-2-7}, i.e., invariant under the orientation-preserving diffeomorphisms of $M$. An infinitesimal variation of \eqref{subsec2-2-11} is
\begin{equation}\label{subsec2-2-12}
    \delta \mathcal{S}=\int_{U}\left(E_a\wedge\delta e^a+E^b{}_a\wedge\delta \omega^a{}_b\right)+\int_{\partial U}\Theta,
\end{equation}
thus the Euler-Lagrange expressions $E^a{}_b$ for $\omega^a{}_b$ are antisymmetric.

The \emph{Noether identity for local Lorentz invariance} of \eqref{subsec2-2-11} is obtained similarly as for metric-affine gauge theory. The only difference is that now $\lambda^a{}_b(x)$ are antisymmetric, i.e., $\lambda_{ab}(x)=\lambda_{[ab]}(x)$, since the local Lorentz $so(1,3)$ Lie algebra consists of $so(1,3)$-valued functions. The Noether identity is 
\begin{equation}\label{subsec2-2-13}
 \mathcal{D}E^a{}_b-E_{[b}\wedge e^{a]}=0.
\end{equation}

The \emph{Noether identity for diffeomorphism invariance} of \eqref{subsec2-2-11} is obtained similarly as for metric-affine gauge theory and reads
\begin{equation}\label{subsec2-2-14}
 \mathcal{D}E_a+E_c\wedge i_aT^c+E^b{}_c\wedge i_a\Omega^c{}_b =0.
\end{equation}

Lorentz gauge theories can be formally obtained as a special case of metric-affine gauge theories. An element of the local $gl(4,\mathbb{R})$ Lie algebra decomposes to its symmetric and antisymmetric parts as $\lambda_{ab}(x)=\lambda_{(ab)}(x)+\lambda_{[ab]}(x)$, where the antisymmetric part corresponds to local Lorentz $so(1,3)$ Lie algebra. If in the set-up of metric-affine gauge theories we first fix $g_{ab}$ as $\eta_{ab}$, then only the local Lorentz symmetry preserves this fixing. This means that the whole local $GL(4,\mathbb{R})$ symmetry breaks down to local Lorentz symmetry\footnote{The described process has a natural interpretation in the language of principal $G$-bundles, which then allows the metric tensor to be treated as a classical Higgs field \cite{Trautman:1979cq, Trautman:1981fd, Ivanenko:1987ka, Sardanashvily:1994qd}.}. To obtain the Lorentz gauge theories, one also requires the vanishing of the non-metricity of the linear connection and the action of the form \eqref{subsec2-2-11}. 

An example of a Lorentz gauge theory is the \emph{Einstein-Cartan} gravitational theory. The source-free action of the Einstein-Cartan theory (also called Palatini version of general relativity) is
\begin{equation}\label{subsec2-2-15}
   \mathcal{S}[e^a,\omega^a{}_b]:= -\frac{1}{2}\int_U \varepsilon_{abcd}\Omega^{ab}\wedge e^c\wedge e^d,  
\end{equation}
where $\Omega^{ab}$ are the curvature 2-forms of a metric connection $\nabla_m$. The Euler-Lagrange expressions are
\begin{equation}\label{subsec2-2-16}
    E_a=\varepsilon_{abcd}\Omega^{bc}\wedge e^d, \qquad E^a{}_b=\eta^{ac}\varepsilon_{cbdf}T^d\wedge e^f. 
\end{equation}

\section{Teleparallel equivalent of general relativity}\label{sec3}
\subsection{General relativity and its teleparallel equivalent}\label{subsec3-2}

\emph{General Relativity} (GR) is a classical gauge gravitational theory geometrically defined by $(M,g,\nabla_{LC})$, where $M$ is a smooth 4-dim.~right-oriented manifold and $\nabla_{LC}$ is the Levi-Civita connection, i.e., the torsion-free and non-metricity-free linear connection. The source-free action (in units $16\pi G/c^4=1$) is\footnote{In this paper, we assume that all variations have compact support in $U$, which avoids the problem of imposing variations of both the fields and their normal derivatives to vanish at the boundary \cite{Poisson:2009pwt}. If we consider variations that vanish on the boundary $\partial U$, we must add the Gibbons-Hawking-York term \cite{York:1972sj, Gibbons:1976ue}. Note that the problem persists even in the Palatini method, where we treat $e^a$ and $\omega^a{}_b$ as independent fields, since in the absence of spin matter the variation with respect to $\omega^a{}_b$ leads to \eqref{subsec3-2-2}, which forces both $\delta e^a$ and its normal derivative to vanish at the boundary \cite{Obukhov1987}.}

\begin{equation}\label{subsec3-2-1}
   \mathring{\mathcal{S}}[e^a]:= -\frac{1}{2}\int_U \varepsilon_{abcd}\mathring{\Omega}^{ab}\wedge e^c\wedge e^d,  
\end{equation}
where $\mathring{\Omega}^{ab}$ are the curvature 2-forms of the Levi-Civita connection 1-forms $\mathring{\omega}^a{}_b=\eta^{ac}\mathring{\omega}_{cb}$ given as
\begin{equation}\label{subsec3-2-2}
    \mathring{\omega}_{ab}=\frac{1}{2}\left(\eta_{ac}i_bde^c-\eta_{bc}i_ade^c+ \eta_{dc}\left(i_ai_bde^d\right)e^c\right).
\end{equation}
The Euler-Lagrange expression of \eqref{subsec3-2-1} is
\begin{equation}\label{subsec3-2-3}
    \mathring{E}_a=\varepsilon_{abcd}\mathring{\Omega}^{bc}\wedge e^d.
\end{equation}
The gauge symmetries of the action \eqref{subsec3-2-1} are the local Lorentz gauge symmetry and the diffeomorphism gauge symmetry (the orientation-preserving diffeomorphisms of $M$), which yield  Noether identities
\begin{equation}\label{subsec3-2-4}
    \mathring{E}_{[a}\wedge e^{b]}=0,\qquad\mathring{\mathcal{D}}\mathring{E}_a=0.
\end{equation}
The first identity tells us that the field equations are symmetric. The second identity, i.e., the Noether identity for the diffeomorphism gauge symmetry, is equivalent to the contracted Bianchi identity, i.e., the contraction of the second Bianchi identity $\mathring{\mathcal{D}}\mathring{\Omega}^a{}_b=0$.

The \emph{Teleparallel Equivalent of General Relativity} (TEGR), also known as \emph{teleparallel gravity}, is a classical gauge gravitational theory geometrically defined as $\left(M,g,\nabla_{T}\right)$, where $M$ is a smooth 4-dimensional~right-oriented manifold and $\nabla_{T}$ is a \emph{teleparallel connection}, i.e., a metric-compatible and curvature-free linear connection, with the teleparallel connection 1-forms given by 
\begin{equation}\label{spinconn}
\dot{\omega}^a{}_b=(\Lambda^{-1})^a{}_c(x) d \Lambda^c{}_b(x),
\end{equation}
where $\Lambda^a{}_b\in SO(1,3)$, and $\dot{\omega}^a{}_b$ are related to the Levi-Civita connection 1-forms \eqref{subsec3-2-2} as
\begin{equation}\label{decomposition}
\dot{\omega}^a{}_b=\mathring{\omega}^a{}_b+\dot{K}^a{}_b,
\end{equation}
where $\dot{K}^a{}_b$ are the contortion 1-forms related to the teleparallel torsion 2-forms as $\dot{T}^a=\dot{K}^a{}_b\wedge e^b$.

The source-free action of TEGR is \cite{Hehl:1994ue,Obukhov:2002tm}
\begin{equation}\label{subsec3-2-5}
    \mathcal{S}[e^a,\dot{\omega}^a{}_b]:= \frac{1}{2}\int_U \dot{T}^a\wedge H_a,
\end{equation}
where $H_a$ is the excitation form defined as
\begin{equation}\label{subsec3-2-6}
     H^a:=* \left(\dot{T}^a-2e^a\wedge i_b \dot{T}^b+e^b\wedge i^a \dot{T}_b\right),
\end{equation}
where $i^a:=\eta^{ab}i_b$ and $*=*_{g(e)}$ is the Hodge star operator.

The Euler-Lagrange expressions of \eqref{subsec3-2-5} are
\begin{equation}\label{subsec3-2-7}
    \dot{E}_a=-\dot{\mathcal{D}}H_a+\mathcal{J}_a,\qquad\dot{E}^a{}_b=0,
\end{equation}
where $\mathcal{J}_a=\frac{1}{2}\left(\dot{T}^b \wedge i_a H_b -i_a \dot{T}^b\wedge H_b\right)$ is the gravitational energy-momentum 3-form. As was pointed out in \cite{Golovnev:2017dox}, the variation with respect to the teleparallel connection must be performed with respect to \eqref{spinconn}, and hence restricted to the form $\delta\dot{\omega}^a{}_b=\dot{\mathcal{D}}\lambda^a{}_b(x)$, where $\lambda^a{}_b(x)$ are antisymmetric.

The gauge symmetries of TEGR \eqref{subsec3-2-5} are local Lorentz transformations and the orientation-preserving diffeomorphisms of $M$. Moreover, TEGR belongs to the Lorentz gauge theory class, and hence Noether identities can be directly obtained from \eqref{subsec2-2-13}, \eqref{subsec2-2-14}, and \eqref{subsec3-2-7}. They read
\begin{equation}\label{subsec3-2-8}
    \dot{E}_{[a}\wedge e^{b]}=0,\qquad \dot{\mathcal{D}}\dot{E}_a+\dot{E}_c\wedge i_a\dot{T}^c=0.
\end{equation}

Let us now focus on the relation between general relativity and TEGR, which explains the logic behind TEGR's name. Using the decomposition \eqref{decomposition}, one can prove the following identity
\begin{equation}\label{subsec3-2-9}
\frac{1}{2}\varepsilon_{abcd}\mathring{\Omega}^{ab}\wedge e^c\wedge e^d=-\frac{1}{2}\dot{T}^a\wedge H_a-2 d(e^a \wedge *\dot{T}_a).
\end{equation}
This identity implies that TEGR is locally dynamically equivalent to GR, i.e., that the tetrad field equations of both theories are the same, i.e., $\mathring{E}_a=\dot{E}_a$, since both actions differ by a boundary term. In TEGR, the field equations for the teleparallel connection are trivial since the Euler-Lagrange expressions $\dot{E}^a{}_b$ vanish identically \eqref{subsec3-2-8}. This means that all teleparallel connections $\nabla_{T}$ trivially satisfy the field equations, and hence $\nabla_{T}$ can be chosen arbitrarily as far as the field equations are concerned\footnote{The teleparallel connection can be determined by imposing some additional physical requirements, such as finiteness of the (Euclidean) action \cite{Krssak:2015rqa,Krssak:2015lba,Krssak:2026jvv},  or vanishing energy-momentum current \cite{Koivisto:2019ejt}.}. The coframe fields $e^a$ are given as the solutions of the field equations and are the same as in general relativity.

Lastly, we want to remark that the actions of general relativity \eqref{subsec3-2-1} and TEGR \eqref{subsec3-2-5} have the same symmetries, since they differ by a boundary term \eqref{subsec3-2-9}. For gauge symmetries, this implies that the Noether identities in TEGR \eqref{subsec3-2-8} are equivalent to those in general relativity \eqref{subsec3-2-4}. The equivalence of Noether identities for local Lorentz symmetry follows directly from $\mathring{E}_a=\dot{E}_a$. To see the equivalence of Noether identities for diffeomorphisms, we use the definition of the contortion, which gives $i_a\dot{T}^c=\left(i_a\dot{K}^c{}_b\right)e^b-\dot{K}^c{}_a$. Inserting this expressions to the Noether identity for diffeomorphisms \eqref{subsec3-2-8}, we obtain
\begin{equation}
\dot{\mathcal{D}}\dot{E}_a+\left(i_a \dot{K}^c{}_b\right) \dot{E}_c\wedge e^b-\dot{E}_c\wedge\dot{K}^c{}_a=0.
\end{equation}
The second term vanishes due to Noether identities for the local Lorentz symmetry and the antisymmetry of the contortion $\dot{K}_{cb}=\dot{K}_{[cb]}$. Using \eqref{decomposition}, the remaining terms can be written as $\dot{\mathcal{D}}\dot{E}_a+\dot{K}^c{}_a\wedge \dot{E}_c=\mathring{\mathcal{D}}\dot{E}_a$. Since $\mathring{E}_a=\dot{E}_a$, this establishes the equivalence of the Noether identities for diffeomorphisms.

\subsection{TEGR and a classical gauge theory of translations}\label{subsec3-3}

In this section, we analyze the presence of local translational gauge symmetry among the gauge symmetries of TEGR from the viewpoint of Noether's second theorem. This approach
sheds light on the gauge structure of TEGR and allows us to understand claims about TEGR being interpreted as a gauge theory of translations.

Let us start by clarifying the term \emph{"local translations"} since this term is intensively used in gauge gravity literature, sometimes even in different non-equivalent meanings. The \emph{group of translations} is given as $T(4):=\left(\mathbb{R}^4,+\right)$, and hence the group operation is addition. We consider the \emph{group of local translations}, which we also call the \emph{local $T(4)$ group}, to be defined as the group of all (smooth) maps $S:U\rightarrow T(4)$ that assign to a point in $U\subset M$ an element of $T(4)$ group. 
This definition is motivated by the situation in gauge theories of internal symmetries, such as electromagnetism, where local $U(1)$ symmetry is a ``local version" of the structure group of a principal $U(1)$-bundle. Therefore, within this definition, local translations are treated as internal gauge transformations.

For TEGR to be considered a gauge theory of translations, the action of TEGR \eqref{subsec3-2-5} has to be invariant under local $T(4)$ transformations, i.e., under local translations defined above, which have to induce infinitesimal variations of $e^a$ and $\dot{\omega}^a{}_b$ of the form \eqref{sec1-7} and lead to a Noether identity. However, this leads to (at least) the following consistency problems:

\begin{enumerate}
    \item In  Section \ref{subsec3-2}, we discussed gauge symmetries of TEGR -- local Lorentz transformations and orientation-preserving diffeomorphisms of $M$. This immediately implies that even if \eqref{subsec3-2-5} were additionally invariant under local translations, TEGR could not be understood solely as a gauge theory of translations due to the presence of local Lorentz and diffeomorphism gauge symmetries.
    \item The second problem is related to how local translations actually act on $e^a$ and $\dot{\omega}^a{}_b$ fields. In our opinion, there generally is no canonical action of local translations on $e^a$ and $\dot{\omega}^a{}_b$ in comparison to how local Lorentz transformations \eqref{subsec2-2-5} and diffeomorphisms \eqref{subsec2-2-10} act. On the other hand, in the gauge gravity literature, the following identifications and suggestions are considered:
    \begin{enumerate}
           \item Most commonly, see for example \cite{Cho:1975dh,Aldrovandi:2013wha,Pereira:2019woq}, the identification $e^a=h^a$ occurs, where $e^a$ is an orthonormal coframe field and $h^a$ are translational connection 1-forms (translational gauge field) obtained as the pullback by a local section of a translational connection 1-form on the principal $T(4)$-bundle. This identification is problematic since $h^a$ are not guaranteed to be non-degenerate, while $e^a$ have to be non-degenerate by definition of a basis of a vector space.
           
            \item Another often made identification, see for example \cite{Pereira:2019woq, Krssak:2018ywd, Koivisto:2019ejt}, is $e^a=B^a+\dot{\mathcal{D}}x^a$, where $e^a$ is an orthonormal coframe field, $B^a$ is the translational part of a Poincaré connection on a Poincaré bundle pullbacked to $U\subset M$, see footnotes \ref{footnote1} and \ref{footnote2}, and $x^a$ are $\mathbb{R}^4$-valued functions in $U$ obtained as the pullback of "tangent space" coordinates on the Poincaré bundle by its local section. The suggested identification is problematic since the right-hand side corresponds to $\Tilde{\mathcal{D}}x^a$, where $\Tilde{\mathcal{D}}$ is the exterior covariant derivative with respect to the pullback of the Poincaré connection to $U$. Thus, $\Tilde{\mathcal{D}}x^a$ is not guaranteed to be non-vanishing, and then the left-hand side cannot be interpreted as an orthonormal coframe field. For more details about this identification, see part 3.3.~in \cite{Brezina:2025dbc}, the relation $(3.8)$ there and the text after it\footnote{Here we take the opportunity to correct a statement in the 4.~paragraph of \cite[p.~12]{Brezina:2025dbc} where is written: "...since $\theta$ is not a connection 1-form on $LM$...", the correct statement is: "...since the translational part of an affine connection is not $ga(4,\mathbb{R})$-valued as a connection on $AM$ has to be by definition...".}. 
    \end{enumerate}
    \item  Noether identities present in TEGR are only the identities for local Lorentz gauge symmetry and the diffeomorphism gauge symmetry \eqref{subsec3-2-8}, i.e., there is no Noether identity for local translational gauge symmetry. This fact then implies that even if the local translational gauge symmetry were present, it would not be an independent gauge symmetry from the local Lorentz and the diffeomorphism gauge symmetries.
\end{enumerate}
We now construct a model of what could be understood under the term "classical gauge theory of translations" to demonstrate how local translational gauge symmetry works. Let us consider a \emph{principal $T(4)$-bundle} over $M$. From the topological point of view, it turns out that this bundle is trivial \cite{LeDelliou:2019esi}, i.e., the total space $P$ is given as $M\times T(4)$. Due to the triviality of the principal $T(4)$-bundle, its group of automorphisms is given as $\text{Aut}~P\approx\text{Aut}_0P\rtimes\text{Diff}~M$, where the group of vertical automorphisms $\text{Aut}_0P$ is isomorphic to the group $\left\{\forall~S:M\rightarrow T(4)\right\}$, where the group operation is addition. Let now the principal $T(4)$-bundle be endowed with a \emph{translational connection} 1-form $\omega^T$. The structure group $T(4)$ is abelian, therefore the curvature 2-form $\Omega^T$ of $\omega^T$ has the form $\Omega^T=d\omega^T$. Let $\sigma_T$ be a local section of the principal $T(4)$-bundle -- due to the triviality of this bundle, we could also consider global sections. Employing this local section, we obtain the \emph{translational gauge field} $h^a:=\sigma_T^*\left(\omega^T\right)^a$ and its \emph{field strength}  $\mathcal{H}^a:=\sigma_T^*\left(\Omega^T\right)^a=dh^a$ in $U$. Let us now consider an action $ \mathcal{S}[h^a]$ that is invariant under local translations, i.e., under the group of smooth maps $\left\{\forall~S:U\rightarrow T(4)\right\}$. Then the infinitesimal action of $S(x)$ on gauge fields $h^a$ induces infinitesimal variations $\delta_{\lambda} h^a=d\lambda^a(x)$, and the corresponding Noether identity for local translational gauge symmetry is
\begin{equation}\label{subsec3-3-4}
dE_a=0.
\end{equation}
Let us assume the existence of an $Ad$-invariant scalar product $\kappa_{ab}$ in the Lie algebra of $T(4)$ group and that the base manifold $M$ is orientable and endowed with a metric tensor field $g$ of signature $(+~-~-~-)$. Then, in this model of a classical gauge theory of translations, we can construct the most general action quadratic in field strength and invariant under local translations as (where $\alpha$ and $\beta$ are constants)
\begin{equation}\label{subsec3-3-1}
    \mathcal{S}[h^a]:=-\frac{1}{2}\int_{U}\left(\alpha\kappa_{ab}\mathcal{H}^a\wedge*\mathcal{H}^b+\beta\kappa_{ab}\mathcal{H}^a\wedge\mathcal{H}^b\right).
\end{equation}
The corresponding Euler-Lagrange expression is
\begin{equation}\label{subsec3-3-2}
    E_a=\alpha\kappa_{ab}d*\mathcal{H}^b.
\end{equation}

Based on the whole analysis, we see that the interpretation of TEGR as a gauge theory of translations is problematic since, according to points 1, 2, and 3, the presence of local Lorentz gauge symmetry and the diffeomorphism gauge symmetry is ignored, problematic identifications of orthonormal (co)frame fields with other objects occur, and the crucial presence of the Noether identity \eqref{subsec3-3-4} for local translations is absent. Moreover, the action of TEGR \eqref{subsec3-2-5} differs significantly from that of the model of a classical gauge theory of translations \eqref{subsec3-3-1}, which is more similar to the actions of electromagnetism \eqref{subsec2-1-1} and Yang-Mills theory \eqref{subsec2-1-6}. On the other hand, the very first proposed teleparallel model by Einstein \cite{Einstein1928} had the form \eqref{subsec3-2-5} containing just the first term in \eqref{subsec3-2-6}, which when expressed through a covariantly constant frame field $\Tilde{e}_a$ \cite[p.~425]{Fecko:2006zy} (also known as parallel frame field or Weitzenböck gauge) obtains the form
\begin{equation}\label{subsec3-3-3}
      \mathcal{S}[\Tilde{e}^a]=\frac{1}{2}\int_{U}\eta_{ab}d\Tilde{e}^a\wedge*d\Tilde{e}^b. 
\end{equation}
The action \eqref{subsec3-3-3} is visually similar with the first term in the action \eqref{subsec3-3-1}, however, these two actions are fundamentally different since the fields $h^a$, $\Tilde{e}^a$ are related to connections defined on different principal $G$-bundles, namely the principal $T(4)$-bundle and the frame bundle $LM$ of $M$. Moreover, in the model of a classical gauge theory of translations \eqref{subsec3-3-1}, it is not even possible to introduce the torsion 2-form, a crucial object in TEGR and gravity in general, since it only exists for linear connections on $LM$ and is closely related to the existence of the canonical 1-form $\theta$ on $LM$. It seems to us that the absence of the concept of torsion in the model of a classical gauge theory of translations is one of the reasons behind the problematic identification $e^a=h^a$ in TEGR \cite{Cho:1975dh,Aldrovandi:2013wha,Pereira:2019woq,Krssak:2018ywd}, which then falsely assigns torsion to $h^a$ and seemingly interprets TEGR as a gauge theory of translations.

We note that in the literature on gauge gravity, for example \cite{Hehl:1994ue, Blagojevic:2003cg, Blagojevic:2012bc, Hehl:2023khc}, it is often claimed that local translations also act on the points of the spacetime manifold $M$, not just on the fields, as we assumed. However, such an interpretation of local translations is not consistent with the definition of local translations as maps $S:U\rightarrow T(4)$ that we used above, and it has nothing to do with a principal $T(4)$-bundle or with the group $T(4)$. What happens is that the term local translation is used as an alternative name for diffeomorphisms of $M$. Indeed, this can be directly seen even from the structure of the automorphism group of the classical gauge theory of translations \eqref{subsec3-3-1} itself - the automorphism group of the principal $T(4)$-bundle is $\text{Aut}~P\approx\left\{\forall~S:M\rightarrow T(4)\right\}\rtimes\text{Diff}~M$, where the groups of local translations $\left\{\forall~S:M\rightarrow T(4)\right\}$ and the diffeomorphism group $\text{Diff}~M$ are explicitly distinguished.
Thus, when the term local translation is used just as another name for a diffeomorphism, then TEGR and all metric-affine and Lorentz gauge gravitational theories are trivially gauge theories with local translations present among their gauge symmetries since they all have a Noether identity for the fundamental diffeomorphism gauge symmetry.

\subsection{Matter coupling in TEGR}\label{subsec3-4}
In this section, we analyze the situation with matter couplings and spinor fields in TEGR. We use Noether's second theorem to explain which matter coupling has to be used and whether spinor fields can be described consistently in TEGR.

Let us consider generic matter actions $\dot{\mathcal{S}}_M$ with the \emph{teleparallel coupling} and $\mathring{\mathcal{S}}_M$ with the \emph{Levi-Civita coupling} that are locally Lorentz invariant and diffeomorphism invariant. The Lagrangian 4-form of the matter action  $\dot{\mathcal{S}}_M\left[e^a,\dot{\omega}^a{}_b,\Psi^i\right]$ depends tensorially on $\left\{e^a,de^a,\dot{\omega}^a{}_b,d\dot{\omega}^a{}_b,\Psi^i,d\Psi^i \right\}$, where $\dot{\omega}^a{}_b$ are teleparallel connection 1-forms and $\Psi^i$ are differential forms of degree $p$ representing matter fields. The Lagrangian 4-form of the matter action $\mathring{\mathcal{S}}_M\left[e^a,\Psi^i\right]$ depends tensorially on $\left\{e^a,de^a,\Psi^i,d\Psi^i \right\}$. This means that for matter actions $\dot{\mathcal{S}}_M$, $\mathring{\mathcal{S}}_M$ we do not assume minimal coupling; however, the assumption of minimal coupling does not change anything in what follows. Infinitesimal compactly supported variations of fields in $U$ yields variations of $\dot{\mathcal{S}}_M$ and $\mathring{\mathcal{S}}_M$ 
\begin{equation}\label{subsec3-4-1}
     \delta\dot{\mathcal{S}}_M=\int_U\left(\dot{M}_a\wedge\delta e^a+M^b{}_a\wedge \delta\dot{\omega}^a{}_b+\dot{\mathcal{E}}_i\wedge\delta\Psi^i\right)), \qquad \delta\mathring{\mathcal{S}}_M=\int_U\left(\mathring{M}_a\wedge\delta e^a+\mathring{\mathcal{E}}_i\wedge\delta\Psi^i\right).
\end{equation}
Local Lorentz and diffeomorphism invariance yield for $\dot{\mathcal{S}}_M$ the matter on-shell Noether identities 
\begin{equation}\label{subsec3-4-2}
     \dot{\mathcal{D}}M^a{}_b-\dot{M}_{[b}\wedge e^{a]} \overset{m}{\approx}~0,\qquad \dot{\mathcal{D}}\dot{M}_a+\dot{M}_c\wedge i_aT^c\overset{m}{\approx}0,
\end{equation}
and for $\mathring{\mathcal{S}}_M$ the matter on-shell Noether identities 
\begin{equation}\label{subsec3-4-3}
    \mathring{M}_{[b}\wedge e^{a]} \overset{m}{\approx}0, \qquad \mathring{\mathcal{D}}\mathring{M}_a\overset{m}{\approx}0,
\end{equation}
where the symbol $\overset{m}{\approx}$ means that matter field equations hold.

Let us consider generic gravity actions $\dot{\mathcal{S}}_G\left[e^a,\dot{\omega}^a{}_b\right]$ and $\mathring{\mathcal{S}}_G\left[e^a\right]$ based on teleparallel and Riemannian geometries. Compactly supported infinitesimal variations of fields in $U$ yields variations of $\mathcal{S}_G$ and $\mathring{\mathcal{S}}_G$ 
\begin{equation}\label{subsec3-4-4}
     \delta\dot{\mathcal{S}}_G=\int_U\left(\dot{E}_a\wedge\delta e^a+E^b{}_a\wedge \delta\dot{\omega}^a{}_b\right), \qquad\delta\mathring{\mathcal{S}}_G=\int_U\left(\mathring{E}_a\wedge\delta e^a\right).
\end{equation}
Local Lorentz and diffeomorphism invariance yield for $\dot{\mathcal{S}}_G$ the off-shell Noether identities 
\begin{equation}\label{subsec3-4-5}
     \dot{\mathcal{D}}E^a{}_b-\dot{E}_{[b}\wedge e^{a]}=~0, \qquad \dot{\mathcal{D}}\dot{E}_a+\dot{E}_c\wedge i_aT^c=0.
\end{equation}
and for $\mathring{\mathcal{S}}_G$ the off-shell Noether identities 
\begin{equation}\label{subsec3-4-6}
    \mathring{E}_{[b}\wedge e^{a]}=0, \qquad \mathring{\mathcal{D}}\mathring{E}_a=0.
\end{equation}

Let actions $\dot{\mathcal{S}}_M$ and $\mathring{\mathcal{S}}_M$ be matter sources for $\dot{\mathcal{S}}_G$ and $\mathring{\mathcal{S}}_M$ for $\mathring{\mathcal{S}}_G$. Compactly supported infinitesimal variations of fields $e^a$ and $\dot{\omega}^a{}_b$ in $U$ yields variations of $\dot{\mathcal{S}}_G+\dot{\mathcal{S}}_M$, $\dot{\mathcal{S}}_G+\mathring{\mathcal{S}}_M$, and $\mathring{\mathcal{S}}_G+\mathring{\mathcal{S}}_M$ 
\begin{subequations}\label{subsec3-4-7}
\begin{align}
  \delta\left(\dot{\mathcal{S}}_G+\dot{\mathcal{S}}_M\right)&=\int_U\left(\left(\dot{E}_a+\dot{M}_a\right)\wedge\delta e^a+\left(E^b{}_a+M^b{}_a\right)\wedge \delta\dot{\omega}^a{}_b\right), \label{subsec3-4-7a}\\
    \delta\left(\dot{\mathcal{S}}_G+\mathring{\mathcal{S}}_M\right)&=\int_U\left(\left(\dot{E}_a+\mathring{M}_a\right)\wedge\delta e^a+E^b{}_a\wedge \delta\dot{\omega}^a{}_b\right), \label{subsec3-4-7b}\\
    \delta\left(\mathring{\mathcal{S}}_G+\mathring{\mathcal{S}}_M\right)&=\int_U\left(\left(\mathring{E}_a+\mathring{M}_a\right)\wedge\delta e^a\right), \label{subsec3-4-7c}
\end{align}
\end{subequations}
resulting in the gravitational Euler-Lagrange equations
\begin{subequations}\label{subsec3-4-8}
\begin{align}
\dot{E}_a&\overset{g}{\approx}-\dot{M}_a, \qquad \dot{\mathcal{D}}E^a{}_b\overset{g}{\approx}-\dot{\mathcal{D}}M^a{}_b, \label{subsec3-4-8a}\\
    \dot{E}_a&\overset{g}{\approx}-\mathring{M}_a, \qquad \dot{\mathcal{D}}E^a{}_b\overset{g}{\approx}0, \label{subsec3-4-8b}\\
   \mathring{E}_a&\overset{g}{\approx}-\mathring{M}_a, \label{subsec3-4-8c}
\end{align}
\end{subequations}
where the symbol $\overset{g}{\approx}$ means that the gravitational field equations hold, and as mentioned in the paragraph below the equation \eqref{subsec3-2-7}, we used the fact that for a teleparallel connection, the infinitesimal variation of an action with respect to $\dot{\omega}^a{}_b$ is in the form $\delta\dot{\omega}^a{}_b=\dot{\mathcal{D}}\lambda^a{}_b(x)$, where we assume $\lambda ^a{}_b(x)$ to be compactly supported in $U$. 

In generic teleparallel gravitational theories $\dot{\mathcal{S}}_G$, the off-shell Noether identities \eqref{subsec3-4-5} together with the field equations \eqref{subsec3-4-8a} and \eqref{subsec3-4-8b} yield on-shell Noether identities
\begin{subequations}\label{subsec3-4-9}
\begin{align}
 \dot{\mathcal{D}}M^a{}_b-\dot{M}_{[b}\wedge e^{a]}&\overset{g}{\approx}0, \qquad \dot{\mathcal{D}}\dot{M}_a+\dot{M}_c\wedge i_aT^c\overset{g}{\approx}0, \label{subsec3-4-9a}\\
    \mathring{M}_{[b}\wedge e^{a]} &\overset{g}{\approx}0, \qquad  \mathring{\mathcal{D}}\mathring{M}_a\overset{g}{\approx}0, \label{subsec3-4-9b}
\end{align}
\end{subequations}
where the second identity in \eqref{subsec3-4-9b} is just reexpression of $\dot{\mathcal{D}}\mathring{M}_a+\mathring{M}_c\wedge i_aT^c\overset{g}{\approx}0$ according to steps provided in the last paragraph of Section \ref{subsec3-2}. The on-shell Noether identities \eqref{subsec3-4-9a} and \eqref{subsec3-4-9b} are consistent with the matter on-shell Noether identities \eqref{subsec3-4-2} and \eqref{subsec3-4-3}. This then implies that in teleparallel theories, such as new general relativity or $f(T)$ gravities, one can consistently choose between the teleparallel and the Levi-Civita couplings, i.e., one has to postulate the coupling prescription that one wants to use. 

On the other hand, the situation with TEGR $\dot{\mathcal{S}}_G=\dot{\mathcal{S}}$ is very different since the structure of Noether identities in TEGR leads just to the Levi-Civita coupling.
Indeed, in TEGR \eqref{subsec3-2-5}, the Euler-Lagrange expressions are \eqref{subsec3-2-7}, i.e., $\dot{E}^a{}_b=\dot{\mathcal{D}}E^a{}_b=0$ vanish off-shell. As mentioned in the last paragraph of Section \ref{subsec3-2}, the Euler-Lagrange expression $\dot{E}_a$ in TEGR is off-shell equal to the $\mathring{E}_a$ in GR, i.e., $\dot{E}_a=\mathring{E}_a$. These two facts imply that off-shell Noether identities in TEGR \eqref{subsec3-2-8} are off-shell Noether identities in GR \eqref{subsec3-2-4}. In GR  $\mathring{\mathcal{S}}_G=\mathring{\mathcal{S}}$, the on-shell Noether identities yield matter on-shell Noether identities \eqref{subsec3-4-3} -- this follows from \eqref{subsec3-4-6} and \eqref{subsec3-4-8c}. Therefore, the structure of TEGR leads to the Levi-Civita coupling, which does not need to be postulated as is, for example, in new general relativity or $f(T)$ theories, but naturally follows from the structure of TEGR itself. 

If one chose to impose the teleparallel coupling in TEGR, then one would face immediate consistency problems. Indeed, the equations \eqref{subsec3-4-8a} and \eqref{subsec3-4-9a} reduce to
\begin{subequations}\label{subsec3-4-10}
\begin{align}
\dot{E}_a&\overset{g}{\approx}-\dot{M}_a, \qquad 0\overset{g}{\approx}-\dot{\mathcal{D}}M^a{}_b, \label{subsec3-4-10a}\\
   \dot{M}_{[b}\wedge e^{a]}&\overset{g}{\approx}0, \qquad \dot{\mathcal{D}}\dot{M}_a+\dot{M}_c\wedge i_aT^c\overset{g}{\approx}0. \label{subsec3-4-10b}
\end{align}
\end{subequations}
The on-shell Noether identity \eqref{subsec3-4-10b} is not consistent with the matter on-shell Noether identity \eqref{subsec3-4-2} and the second equation in \eqref{subsec3-4-10a} constrains possible $M^a{}_b$ and $\dot{\omega}^a{}_b$. It is often suggested \cite{Maluf:2003fs,Obukhov:2004hv, Aldrovandi:2013wha, BeltranJimenez:2020sih,Hohmann:2022mlc} to consider the Levi-Civita coupling for consistency reasons and the fact that \eqref{subsec3-4-10b} can be reexpressed as 
\begin{equation}\label{subsec3-4-11}
    \mathring{M}_{[b}\wedge e^{a]}\overset{g}{\approx}0, \qquad \mathring{\mathcal{D}}\mathring{M}_a\overset{g}{\approx}0,
\end{equation}
since the off-shell relation $\dot{E}_a=\mathring{E}_a$ between the gravitational Euler-Lagrange expressions in TEGR and GR implies through \eqref{subsec3-4-8c} and the first equation in \eqref{subsec3-4-10a} that $\dot{M}_a\overset{g}{\approx}\mathring{M}_a$. However, as we explained above, the structure of Noether identities in TEGR naturally leads to the Levi-Civita coupling prescription, which can also be seen from reexpressing \eqref{subsec3-4-10b} into \eqref{subsec3-4-11}.

Lastly, we want to comment on the presence of spinor fields in TEGR, since it is often claimed, see, for example, \cite{Obukhov:2002tm,Blagojevic:2012bc,Ferraro:2022xwk}, that spinor fields cannot be consistently described in TEGR. This also allows us to explicitly demonstrate the situation with coupling prescriptions in TEGR, which has been discussed only abstractly so far.

Let the oriented Riemannian manifold endowed with a metric connection admit a spin structure; let $\omega^a{}_b$ be the connection 1-forms of the metric connection, $\Psi$ be a $V$-valued 0-form, where $V$ is the vector space in which the spinor representation acts. We consider\footnote{One could also consider Hermitian Dirac action and mass term; however, this does not affect the following conclusion.}  \emph{Dirac action} of spinor fields $\Psi$ 
\begin{equation}\label{subsec3-4-12}
   \mathcal{S}_M\left[e^a,\omega^a{}_b,\Psi,\bar{\Psi}\right]:=\int_U\left(\frac{1}{6}\varepsilon_{abcd}\left(\bar{\Psi}\gamma^a\mathcal{D}\Psi\right)\wedge e^b\wedge e^c\wedge e^d\right), \quad \mathcal{D}\Psi=d\Psi+\frac{1}{8}\omega_{ab}\left[\gamma^a,\gamma^b\right]\Psi.
\end{equation}
The infinitesimal variation of \eqref{subsec3-4-12} with respect to $e^a$ and the metric connection 1-forms $\omega^a{}_b$ give
\begin{equation}\label{subsec3-4-13}
M_a=-\frac{1}{2}\varepsilon_{abcd}\left(\bar{\Psi}\gamma^b\mathcal{D}\Psi\right)\wedge e^c\wedge e^d, \quad M^a{}_b=-\frac{1}{48}\eta_{bc}\bar{\Psi}\gamma^d\left[\gamma^c,\gamma^a\right]\Psi \varepsilon_{dfgh} e^f\wedge e^g\wedge e^h.
\end{equation}

Let us now assume that the metric connection in \eqref{subsec3-4-12} is also curvature-free. Then the variations of \eqref{subsec3-4-12} with respect to $e^a$ and $\dot{\omega}^a{}_b$ give $\dot{M}_a=M_a$ and $\dot{M}^a{}_b=\dot{\mathcal{D}}M^a{}_b$ as the Euler-Lagrange expressions, where $M_a$ and $M^a{}_b$ have the form \eqref{subsec3-4-13} with the teleparallel $\dot{\mathcal{D}}$. The expression $\dot{\mathcal{D}}M^a{}_b$ becomes a constraint in the TEGR on-shell case due to the second equation in \eqref{subsec3-4-10a}. This constraint on spinor fields and the teleparallel connection arises only from imposing the teleparallel coupling in the Dirac action \eqref{subsec3-4-12} and can generally lead to consistency problems and possible restrictions on $\Psi$ and $\dot{\omega}^a{}_b$. On the other hand, if one imposes the Levi-Civita coupling in \eqref{subsec3-4-12}, i.e., requires the metric connection to be torsion-free instead of curvature-free, then one obtains $\mathring{M}_a$ given as
\begin{equation}\label{subsec3-4-14}
\mathring{M}_a=-\frac{1}{2}\left(\varepsilon_{abcd}\left(\bar{\Psi}\gamma^b\mathring{\mathcal{D}}\Psi\right)-\frac{1}{8}\varepsilon_{fgcd}\mathring{\mathcal{D}}\left(\bar{\Psi}\left(\gamma^f\left[\gamma_a,\gamma^g\right]+\frac{1}{2}\gamma_a\left[\gamma^f,\gamma^g\right]\right)\Psi\right)\right)\wedge e^c\wedge e^d.
\end{equation}
Therefore, the Levi-Civita coupling, which naturally follows from the structure of TEGR, allows one to incorporate spinor fields into TEGR with no constraint nor consistency problems occurring.

\section{Discussion and conclusions}

Noether's second theorem states that invariance of an action under an infinite-dimensional variational symmetry implies that the Euler--Lagrange expressions are not independent but satisfy off-shell differential identities known as Noether identities. These identities encode the redundancy of the dynamical variables, and every genuine local gauge symmetry is associated with a corresponding Noether identity. We have first reviewed Noether's theorems in general and derived the Noether identities in relevant cases, including electromagnetism, Yang-Mills theory, metric-affine and Lorentz gauge theories of gravity. 

We have then applied this analysis to TEGR, showing that the gauge symmetries are (orientation-preserving) diffeomorphisms and local Lorentz transformations, since only these symmetries give rise to Noether identities \eqref{subsec3-2-8}. This can be compared with our recent work \cite{Brezina:2025dbc}, where the gauge structure of TEGR was examined using the framework of Trautman, in which gauge symmetries are defined as automorphisms of a principal bundle preserving absolute elements of the theory, which led us to conclude that the diffeomorphisms are the gauge symmetry of TEGR. 

The apparent disagreement between these two conclusions becomes clearer when we compare the roles of diffeomorphism and local Lorentz transformations in gravity with those of gauge transformations in Yang--Mills-type theories and general relativity. In electromagnetism and Yang--Mills theory, the local gauge symmetries give rise to corresponding divergence-type Noether identities, \eqref{subsec2-1-3} and \eqref{subsec2-1-4}, respectively. The same is true of diffeomorphism invariance in general relativity, which gives rise to the contracted Bianchi identity in both the metric and tetrad formulations. By contrast, the additional local Lorentz symmetry of the tetrad formulation gives rise to an algebraic Noether identity, as shown in \eqref{subsec3-2-4}. This highlights the more fundamental role of diffeomorphism invariance, and supports the suggestion that it plays the role analogous to the gauge symmetry in Yang--Mills theories.

Local Lorentz symmetry, on the other hand, reflects the additional freedom introduced by the tetrad formulation and can be eliminated by expressing the theory in terms of the metric, which is invariant under local Lorentz transformations \cite{Golovnev:2022rui}. In this sense, the symmetry arises merely from the choice of dynamical variables and does not alter physical degrees of freedom of the theory, since it can be factored out by passing to metric variables. Nevertheless, from a geometrical perspective, local Lorentz symmetry is the unavoidable consequence of the underlying principal bundle structure. It is therefore interesting that Trautman's framework captures this subtle distinction and -- despite using the frame bundle and tetrads as dynamical variables -- identifies diffeomorphisms as the relevant gauge symmetry of the theory. 

The Hamiltonian formulation provides another perspective on this distinction.  In Yang-Mills-type of theories, each gauge symmetry leads to both primary and secondary constraints, which are all first class and hence remove two degrees of freedoms, leading to the common saying that ``gauge hits twice" \cite{Henneaux:1992ig,Golovnev:2022rui}. The same pattern occurs in general relativity, where diffeomorphism invariance generates both primary and secondary first-class constraints. The situation is different for local Lorentz symmetry in the tetrad formulation, which generates only primary first-class constraints \cite{Deser:1976ay,Nelson:1978ex,Flinckman:2026kpw}. The fact that the local Lorentz symmetry generates first-class constraints confirms that it should be regarded as a gauge symmetry, but it is an example of a gauge symmetry that fails to hit twice\footnote{We consider here only the second-order formulation of tetrad general relativity. The situation is different in the first-order formulation, where additional constraints can appear \cite{Castellani:1981ue,Duque:2025spv}}. This again shows the different nature of diffeomorphism and local Lorentz symmetries in gravity, and confirms our suggestion that the former plays a role analogous to that of the gauge symmetry in Yang--Mills-type theories. 

Therefore, the gauge structure of TEGR closely resembles that of general relativity in its tetrad formulation. Indeed, the Noether identities \eqref{subsec3-2-4} and \eqref{subsec3-2-8} are equivalent, as shown at the end of Section~\ref{subsec3-2}, which demonstrates that the same gauge symmetries are underlying both theories. The crucial difference between these theories is that the teleparallel connection \eqref{spinconn} is not given in terms of the tetrads but is treated as an independent variable.  The corresponding Noether identities also exhibit an interesting difference in their structure. In Lorentz gauge theory, they involve the Euler--Lagrange expressions associated with both the tetrad and the connection, whereas in TEGR, owing to the identity $\dot{E}^a{}_b=0$, they involve only the Euler--Lagrange expression associated with the tetrad, $\dot{E}_a$. The latter therefore coincide with the Noether identities obtained in the non-covariant formulation of TEGR\footnote{This just shows that the dynamical content of both the covariant and non-covariant formulations of TEGR is fully equivalent.}, as well as with those in the tetrad formulation of general relativity. This result is in agreement with the Hamiltonian analysis of TEGR, and more generally of modified teleparallel theories of gravity, which shows that the independent teleparallel connection does not introduce or remove physical degrees of freedom \cite{Blixt:2018znp,Blixt:2019ene,Blixt:2020ekl,Blixt:2022rpl}.

We have then use the results of our Noether analysis to examine the interpretation of TEGR as a classical gauge theory of translation. We have shown that this interpretation faces several conceptual and technical difficulties. First, it does not adequately account for the diffeomorphism and local Lorentz gauge symmetries already present in TEGR. Second, it relies on problematic identifications of orthonormal (co)frame fields with the translational gauge potentials/connections. Third, the proposed interpretation does not establish the existence of a Noether identity associated with an independent local translational gauge symmetry \eqref{subsec3-3-4}, as would be expected if such a symmetry were fundamental to the theory. To demonstrate the differences, we have constructed also what should be regarded as the gauge theory of translations, given by the action \eqref{subsec3-3-1}, which differs from the TEGR action \eqref{subsec3-2-5}. These differences confirm previous findings regarding the difficulties associated with interpreting TEGR as a translational gauge theory \cite{LeDelliou:2019esi,Brezina:2025dbc} and have implications beyond TEGR, extending not only to the previously explored metric-affine theories of gravity \cite{Dass:1984qk,Sardanashvily:1983xbn,Ivanenko:1983fts,Sardanashvily:1984dx,Francois:2025jro}, but also to various recent pre-geometric approaches to gravity \cite{Koivisto:2019ejt,Addazi:2024rzo}. 

We have also examined the coupling of matter, including spinor fields, in TEGR. We have demonstrated that the structure of its Noether identities -- associated with diffeomorphism and local Lorentz gauge symmetries -- uniquely selects the Levi-Civita coupling prescription, whereas the teleparallel coupling prescription generally leads to inconsistencies. Thus, the Levi-Civita coupling is not an additional assumption in TEGR, but follows directly from the gauge structure of the theory. In modified teleparallel gravity theories, such as New General Relativity and $f(T)$ gravity, by contrast, both the Levi-Civita and teleparallel coupling prescriptions can generally be implemented consistently, leaving the choice between them to be specified. In TEGR, however, the only consistent possibility is the Levi-Civita coupling, as in general relativity. Furthermore, this prescription allows spinor fields to be incorporated into TEGR without introducing additional constraints or consistency problems, contrary to claims often found in the literature. These results establish that TEGR and GR are locally on-shell equivalent not only in vacuum, as is well known, but also in the presence of arbitrary matter fields.

\section*{Acknowledgements} 
 This work was supported by VEGA grant 1/0565/25.
 
\bibliography{references}\addcontentsline{toc}{section}{References}
\bibliographystyle{Style}

\end{document}